\documentclass[reprint]{revtex4-2}

\usepackage[english]{babel}
\usepackage{gensymb}
\usepackage{graphicx} 
\usepackage{amsmath}

\begin{document}


\title{Lagrangian Particle Tracking at Large Reynolds Numbers} 



\author{Christian K\"uchler}
\affiliation{Max-Planck-Institute for Dynamics and Self-Organisation, G\"ottingen, Germany}
\author{Antonio Ibanez Landeta}
\affiliation{Max-Planck-Institute for Dynamics and Self-Organisation, G\"ottingen, Germany}
\author{Jan Molacek}
\affiliation{Max-Planck-Institute for Dynamics and Self-Organisation, G\"ottingen, Germany}
\author{Eberhard Bodenschatz}
\affiliation{Max-Planck-Institute for Dynamics and Self-Organisation, G\"ottingen, Germany}
\affiliation{Cornell University, Ithaca, NY, USA}
\email{eberhard.bodenschatz@ds.mpg.de}

\date{\today}

\begin{abstract}

Particle tracking in turbulent flows is fundamental to the study of the transport of tracers, inertial particles or even active objects in space and time, i.e. the Lagrangian frame of reference. It provides experimental tests of theoretical predictions\cite{Toschi2009} (e.g. for the statistics of fluid accelerations and particle dispersion) and helps to understand important natural processes where particle inertia is important (e.g. cloud microphysics\cite{Bertens2021}). While the spatial (Eulerian) properties of turbulent flows have been studied for high, atmospheric Reynolds numbers\cite{Tsuji2004} ($R_\lambda > 10^4$), the profound difficulties in accurately tracking particles in turbulent flows have limited the Reynolds numbers in the Lagrangian reference frame to the Taylor scale Reynolds numbers $R_\lambda \lesssim 10^3$. Here we describe a setup that allowed Lagrangian particle tracking at $R_\lambda$ between 100 and 6000 in the Max Planck Variable Density Turbulence Tunnel (VDTT). We describe the imaging setup within the pressurised facility, the laser illumination, the particles and the particle dispersion mechanism. We verify that the KOBO Cellulobeads D-10 particles are suitable tracers. They carry negligible charge and their Stokes number is small over the full range of experimental conditions. We present typical data from the experiment and discuss the challenges and constraints of the setup. 
\end{abstract}


\maketitle 

\section{Introduction}
Turbulence can be described both in the stationary Eulerian reference frame, i.e. by studying snapshots of vector fields or in the co-moving Lagrangian reference frame of a material element \cite{Monin1975}. While measurements in the Eulerian frame of reference are being performed routinely since the early 1900s \cite{Comte-Bellot1966}, Lagrangian measurements at very high turbulence levels  remain utmost  challenging. In particular, the authors know of no experimental Lagrangian measurements with a Taylor-scale Reynolds number $R_\lambda$ exceeding 1000, even though suitable flows exist in the laboratory \cite{Bodenschatz2014,Rousset2014,Saint-Michel2014,Salort2010} and in the atmosphere \cite{Kahalerras1998,Tsuji2004}. Large Reynolds numbers are desirable, because they are known to reveal universal properties of turbulent flows that are obscured by viscous effects at lower Reynolds numbers. In the Lagrangian reference frame viscous effects diminish slower with increasing $R_\lambda$, making large Reynolds numbers even more relevant than in the Eulerian reference frame \cite{Toschi2009}. 

For 3D incompressible fluid turbulence the statistics of turbulent velocity fluctuations can be  understood in the Eulerian frame by the transfer of  kinetic energy from large to small spatial scales  with  a rate  $\varepsilon$ (power per unit mass). At the small spatial scales this energy is dissipated into heat. $\varepsilon$ is thus called energy dissipation rate. The largest flow length- and time scales ($L$ and $T_L$, respectively) are given by the energy injection mechanism, whereas the fluid viscosity $\nu$ determines the dynamics at the viscous energy dissipating (Kolmogorov) scales $\eta$ and $\tau_\eta$, respectively.  In the Lagrangian frame even for a Taylor-scale Reynolds number $R_\lambda=2000$, which is considered large for the Eulerian frame, the separation in temporal scales is only $T_L/\tau_\eta \approx 300$.  Thus, to this date there is little knowledge on  Lagrangian properties of fully developed turbulence in experiments and idealized numerical simulations  with the most significant one  being acceleration statistics of Lagrangian tracers.\cite{Buaria2022}. 

Lagrangian measurements rest on the tracking of small particles chosen to follow the fluid material elements as passive tracers. From these tracks the local flow velocity $\mathbf{u}$ and acceleration $\mathbf{a}$ can be inferred.
The particles' ability to respond to a given change in the fluid motion is given by its response time \cite{Tropea2007,Voth1998}
\begin{equation}
    \tau_p=\frac{1}{18}d^2\frac{(\rho_P-\rho_F)}{\nu_F \rho_F}, 
    \label{eq:t_p}
\end{equation}
where $d$ is the particle diameter, $\rho_P$ and $\rho_F$ the particle- and fluid densities, respectively, and $\nu_F$ is the kinematic viscosity of the fluid. 
This response time is compared to the viscous time scale $\tau_\eta$ to yield the Stokes number $\mathrm{St}=\tau_p/\tau_\eta$, which characterises the particles' ability to follow a given turbulent flow. In order to relate the particles' velocity and acceleration to those of the flow, the particle size should furthermore not significantly exceed the viscous (Kolmogorov) length scale of the turbulence \cite{Voth2002}. 

Common tracer particles for water flows include polysterene particles \cite{Voth1998}. In gaseous flows droplets of vegetable oil or Glycol-water solutions, particularly small solid particles of TiO\textsubscript{2}\cite{Rong2014}, and helium-filled soap bubbles have been options. The latter are relatively large, and thus easy to visualise, while the helium reduces their mean density close to that of air.  Particles can also be 'designed' to allow for local measurements of flow quantities, such as vorticity \cite{Wu2019} or the entire velocity gradient tensor \cite{Hejazi2019}.

In flow diagnostics, particles are typically  illuminated using high-intensity pulsed laser beams, such that even particles with a very small surface area scatter enough light to be detected by commercial cameras. Note that choosing very small particles to obtain a fast particle response time $\tau_p$ reduces the amount of scattered light and increases the illumination requirements. It is thus advantageous to optimise $\rho_P-\rho_F$ in eq. (\ref{eq:t_p}), e.g. by using larger but lighter particles (with the associated small Stokes number). Particles should not be larger than the Kolmogorov scale $\eta$ of the turbulence as they should not average the flow due to their size. Recently, light-emitting diode arrays have been assembled to deliver less hazardous illumination at reduced costs \cite{Buchmann,Chetelat2002}. Another alternative are fluorescent\cite{Guimaraes2016} or phosphorescent \cite{Kemp2010} particles, which can reduce the demands on the light source intensity at the expense of a very limited selection of particles. 

The imaging device is the central limiting factor in any particle tracking setup, which must first and foremost be fast enough in terms of Kolmogorov times ($>10~ \mathrm{frames}/\tau_\eta$)\cite{Voth2002}. The relatively small amount of light scattered by a single particle furthermore demands a very sensitive sensor. The optical elements must then be chosen such that a single particle image (particle projection and optical aberrations) spans at least 2 pixels to achieve sub-pixel positioning precision \cite{Nobach2005a}. 
The first optical measurements of fully developed Lagrangian turbulence were recorded on silicon-strip detectors from a high-energy physics apparatus \cite{Voth1998,Voth2002,LaPorta2001}. While CCD camera sensors permit Lagrangian particle tracking only at moderate Reynolds numbers\cite{Dracos1996}, the large number of readout channels on CMOS sensors enable frame rates of up to 25 Gpx/s in commercial high-speed cameras \cite{Xu2008,Lawson2018}. The rapid permanent storage of this massive stream of data is time-consuming in practice and a field of active industrial development efforts. 

To obtain long particle tracks, it is advantageous to remove any mean motion of the particles across the limited field of view. Hence, most Lagrangian experiments are performed in flows without a mean flow, such as von-K\'arm\'an mixers \cite{Voth1998,Xu2006,Ouellette2006} or specially designed turbulence generators \cite{Zimmermann2010,Bewley2012}. In wind- and water tunnels, the camera and illumination systems are placed on carriages that move at the mean flow speed \cite{Ayyalasomayajula2006} or the length of the particle tracks is sacrificed in simpler stationary setups \cite{Stelzenmuller2017}. 

\begin{figure*}[hbt]
  \includegraphics[width=\textwidth]{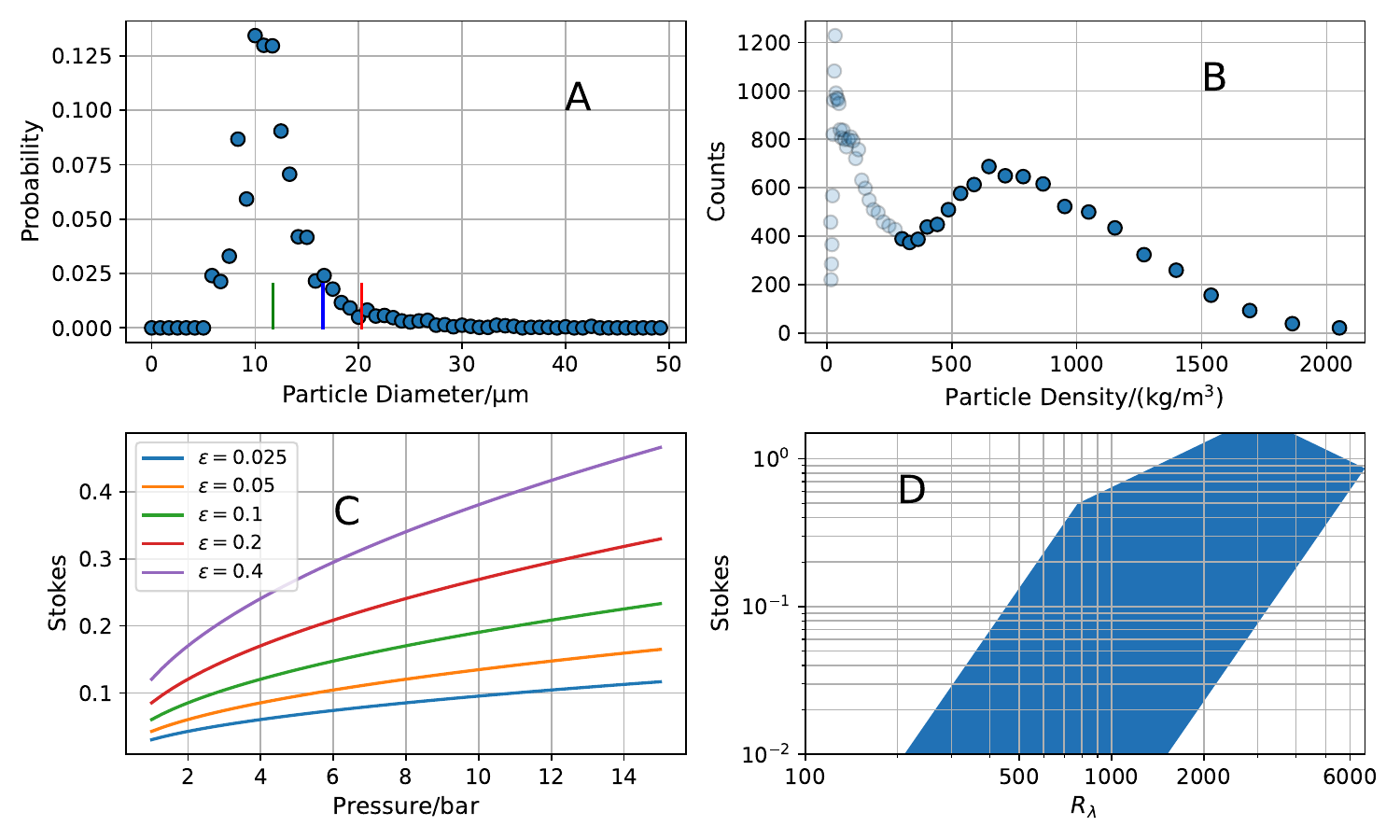}
  \caption{(A): Probability distribution of particle sizes determined from microscopic images of settled particles from typical experiments. Colors indicate expected values for clusters of one (green), two (blue), or three (red) particles. Mean particle diameter in the green sector is $(11.7 \pm 0.1)\mu$m. (B) Histogram of the particle density estimated from the particle response time to a calibrated fluid acceleration. The peak around $2.5 \mu$m is likely a residual from a previous experiment.  From this distribution we estimate the effective particle density and its error as $\rho_p=730$ kg/m$^3$. (C) Estimate of the Stokes number using the values of particle diameter and -density along with known fluid and turbulence properties. (D) Range of Stokes numbers obtainable at given $R_\lambda$. $\mathrm{St}<0.1$ can only be reached for $R_\lambda<3000$.}
  \label{Fig:ParticleSizePlot}
\end{figure*}

A wide range of software packages have been developed that allow the frame-by-frame tracking of imaged particles and the subsequent transformation of two-dimensional pixel coordinates to three-dimensional "world" coordinates. The conventional method is to employ a predictor-corrector scheme \cite{Ouellette2006a,Xu2008,Malik1993} and a triangulation based on the simple pinhole camera model \cite{Tsai1987}. The recent Shake-the-box algorithm \cite{Schanz2016} allows much larger particle densities on the images, makes the particle identification more efficient, and performs better overall. Recently, an open source code based on the shake-the-box algorithm has been made available \cite{Tan2020}.

At large Reynolds numbers, Lagrangian fluctuations occur over a wide range of time scales. Small particles move vigorously on small time scales, while the tracks remain statistically correlated for long times. Capturing these rich motions by imaging is thus a measurement challenge, which needs to be embedded in the substantial experimental creativity required to create large Reynolds numbers in the first place. 

This article introduces a solution to this problem in the form of a stationary Lagrangian particle tracking system in a high-pressure, active-grid driven wind tunnel flow. It is organised as follows: First we introduce the most important aspects of the Variable Density Turbulence Tunnel and mention the advantages and disadvantages of SF\textsubscript{6} as a working gas. We then proceed to characterise the particles of choice and describe the dispersion mechanism. We further introduce the laser illumination and the camera setup. We then detail our in-house particle tracking code and present first measurements to demonstrate the successful operation of the experiment. We conclude by studying the overall capabilities of the setup. 

\section{Facility and Flow Properties}
Generating large Reynolds numbers in approximately homogeneous and isotropic laboratory flows has challenged the scientific community for decades (see above). The Variable Density Turbulence Tunnel (VDTT) \cite{Bodenschatz2014} solves this problem in a particularly flexible manner. It relies on the use of SF\textsubscript{6} at pressures up to 15 bar, which is non-corrosive, non-toxic, but a strong greenhouse gas and emissions must be kept at a minimum. SF\textsubscript{6} is furthermore at equal pressure about $6\times$ denser than air and therefore has a lower kinematic viscosity $\nu$. Since $R_\lambda \sim 1/\nu$, its use is beneficial in the generation of large Reynolds numbers. 

The VDTT is a closed-loop wind tunnel and reaches mean flow speeds between 0.5 and  5.5 m/s. In the following we review the wind tunnel measurement section and kindly refer the reader to the original publication \cite{Bodenschatz2014} for further details on the facility. 

The flow enters the measurement section through an active grid \cite{Griffin2019,Kuchler2019}. The active grid consists of 111 motorised winglets ($11~\mathrm{cm}~\times~11~\mathrm{cm}$), whose angle with respect to the mean flow velocity can be individually adjusted over 180 degrees at a speed of up to 40 degrees per 0.1 seconds. By correlating these angles in space and time, the blockage of the tunnel cross section can be controlled locally and dynamically. It has been shown that the active grid can generate correlated fluid structures of variable size \cite{Griffin2019} while maintaining an adequate shear-free central region \cite{Kuchler2020}. This homogeneous region can be expanded by systematically reducing the active grid blockage of the active grid flaps close to the wall. Most importantly, it allows for a fine control of the length scale where turbulent kinetic energy is injected into the flow and therefore the Reynolds number. It needs to be emphasized here that the variable density aspect of the wind tunnel allows to adjust the Reynolds number not only by adjusting the energy injection scale or the fluctuating velocity, but also the kinematic viscosity.  Bringing together control over the smallest length scales through density (i.e. pressure) and control over the largest scales through the active grid results in the aforementioned parameter flexibility and large yet unsurpassed Reynolds numbers at mean flow speeds of only up to 5m/s. In terms of measurement instrumentation this allows for detailed investigations of finite-resolution effects, since one Reynolds number can be reproduced with different combinations of parameters \cite{Kuchler2019,Kuchler2020,Hutchins2015}. 

After passing through the active grid, the now highly turbulent flow moves through an expansion towards the measurement section height of 117 cm. The height of the measurement section increases slightly further downstream through an inclination of about 0.114\degree~over the entire length of 8.8~m, i.e. about 1.8~cm to avoid the influence of the turbulence generated at the walls of the tunnel on the bulk turbulence. 

It has been shown \cite{Kuchler2020a} that the turbulent kinetic energy decays downstream of the active grid, while its integral length scale defined as $L=\int \langle u(x)\cdot u(x+r) \rangle dr$ remains approximately constant. This indicates that the turbulence decay is influenced by the finite size of the measurement section. 

The particle tracking measurement volume is located 7 m downstream of the active grid, 54-58~cm above the floor and in the center of the 1.5~m wide measurement section. An 8~cm wide traverse is located approximately 1.3~m downstream of the measurement volumes with hot wires protruding about 20~cm. The spacing of the active grid paddles is 12cm and the largest  energy injection scales are 60cm.

\begin{figure*}[t]
  \includegraphics[width=\textwidth]{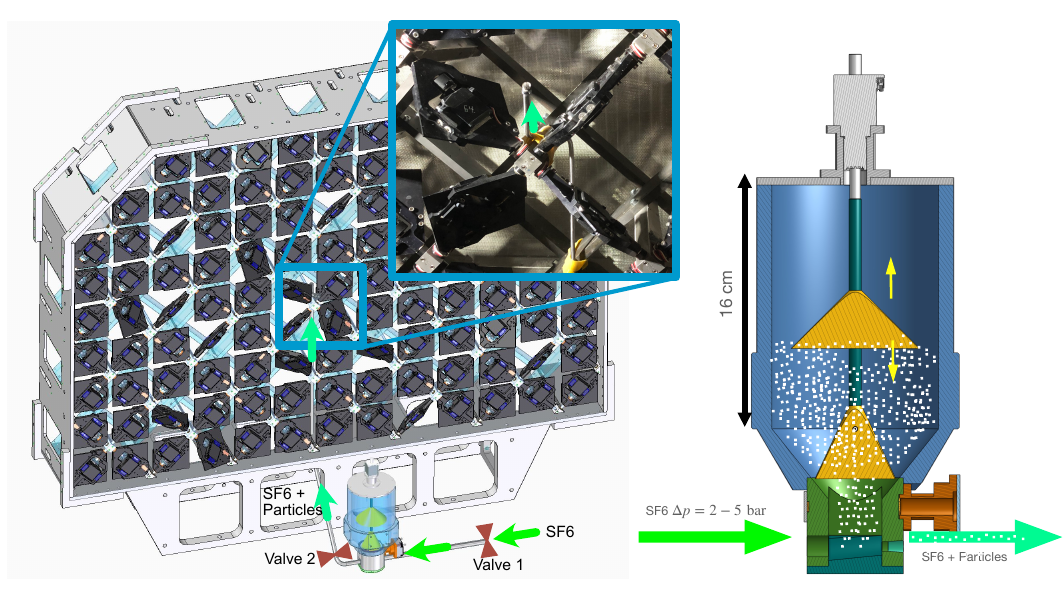}
  \caption{Particle dispersion setup at the active grid. An external SF\textsubscript{6} supply (3-5 bar above facility pressure) is connected to the particle reservoir, where the gas flow collects particles. The reservoir is pressurised upon opening of valve 1 (located outside the wind tunnel), the gas flow is started by opening valve 2. The particles are released through a 0.5mm  aperture nozzle. The particles travel approximately 2~m inside the pipes. The flaps have a side length of 0.11~m.}
  \label{Fig:ParticleDisperser}
\end{figure*}
While being of great benefit in the generation of high Reynolds numbers in a laboratory, SF\textsubscript{6} complicates various aspects of the flow measurement. First, its low kinematic viscosity causes the viscous length- and time scales ($\eta>10 \mathrm{\mu m}$, and $\tau_\eta>0.1~\mathrm{ms}$, respectively) to be an order of magnitude smaller than in air at atmospheric pressures. Consequently, small particles with a fast fluid response are required, and the requirements on spatial and temporal resolution of the imaging are even larger than in air. 
Furthermore, the refractive index $n$ of SF\textsubscript{6} is sensitive towards changes in fluid pressure and temperature\cite{Thomas1988}. For example, an increase in pressure from 10 to 11 bar changes $n$ by 0.008 in SF\textsubscript{6}, but by only 0.0003 in air \cite{Xiao2005}. In SF\textsubscript{6} at 10 bars, a temperature change of 1K leads to a change in the refractive index of SF\textsubscript{6} comparable to that of a 200K increase in air at standard pressure. 

Hence, the line-of-sight of the cameras cannot contain large temperature gradients and all focal lengths must be remote-controllable to account for changes in the facility pressure. The camera calibration must equally be based on a remote-controlled system that can be brought into and removed from the camera's field of view. 

\section{Particles and Particle Dispersion}\label{sec:Particles}
To measure the motion in the turbulent flow, we seed it with KOBO Cellulobeads D-10. These are cellulose particles and typically serve as a primary product for the cosmetics industry, where they are used to diffusively scatter light. They have been designed to be biologically degradable and therefore pose only minimal health risks. This is in contrast to other solid particles used to seed turbulent flows, such as hollow glass spheres, TiO\textsubscript{2}, or polysterene particles. Cellulobeads
are however flammable and their disposal in air demands typical safety precautions for dusty environments. In an SF\textsubscript{6} atmosphere this is not an issue, since the gas is inert.

The quality of a particle as flow tracer is given by the Stokes number $\mathrm{St}=\tau_p/\tau_\eta$. We now detail the methods we used to estimate St. The particle radius $a$ is straightforward to measure using microscopic images of single particles as described below. The fluid density and viscosity is well known \cite{Hoogland1985}. To calculate the particle response time $\tau_p$ the particle density must be measured. For this we repurpose a TSI Aerodynamic Particle Sizer Model 3321 (APS), which directly measures the particle response time through a time-of-flight technique and infers the particle diameter from a pre-defined particle density. The aerodynamic particle size is then given by \cite{DeCarlo2004} 
\begin{equation}
	d_{ae} = 2a \sqrt{F \frac{\rho_p}{\rho_{Ref}}},
\end{equation}
where $\rho_{Ref}$ is a device-specific reference density and $F$ is a shape factor taking into account differences in fluid response for different particle shapes. Since the particles are spherical to a good approximation, their shape factors are taken as $F=1$. Comparing the APS output diameter with the geometric diameter $2a$ yields a measure for the particle density. Fig. \ref{Fig:ParticleSizePlot} (B) shows a distribution of the particle density calculated in this way. The greyed out part of the plot is likely due to a residual from a previous experiment. The relevant part shows a relatively wide distribution of particle densities from which we estimate $\rho_p=(730\pm250)~\mathrm{kg/m^3}$. This amounts to about half the nominal density of cellulose \cite{cellulose_density} indicating that the particles contain voids. 

In combination, the measurement of $a$ and $\rho_p$ allow for an estimate of the Stokes numbers to be expected for different flow parameter settings. Fig. \ref{Fig:ParticleSizePlot} (C) shows the Stokes number for different values of the facility pressure and the turbulence dissipation rate. Only at small pressures or dissipation rates can we expect a Stokes number below 0.1. For this reason, the flexibility of the facility is crucial as it allows the Stokes number to change while keeping the Reynolds number constant to separate effects of St and $R_\lambda$. This is illustrated in Fig. \ref{Fig:ParticleDisperser} (D), where we show the parameter space of St and $R_\lambda$ spanned by the facility when using Cellulobead D-10 particles to seed the flow. In addition, larger Cellulobeads can be used to reach an even wider range in St. The imaging of smaller Cellulobeads would require a more powerful lighting system than used here. 

\begin{figure*}
    \centering
    \includegraphics[width=\textwidth]{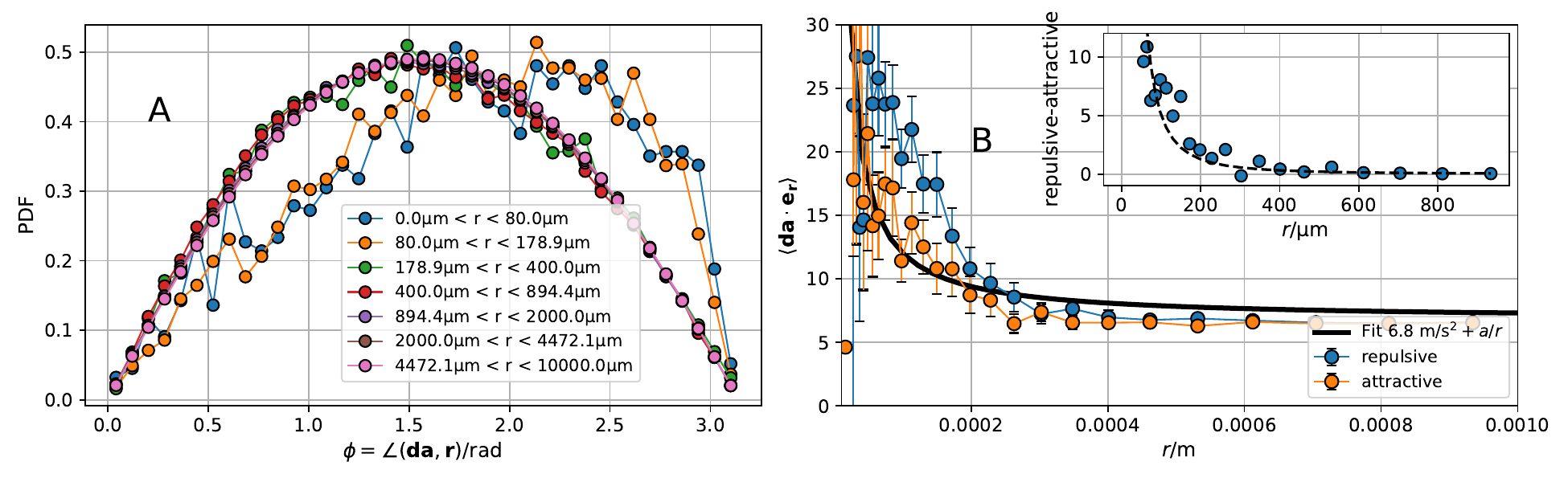}
    \caption{(A): PDFs of the angle between the line of separation of two particles and the net acceleration (i.e. interaction force) between them. Parallel (antiparallel) alignment indicates a attractive (repulsive) interaction, whereas a perpendicular alignment is typical for an incompressible fluid. (B): Acceleration component along the line of separation of two particles. When the component is negative (positive), the interaction is seen as repulsive (attractive). The solid black line is a fit\cite{Borgas1998} $\sim1/r$. \textit{Inset:} Difference between parallel (attractive) and antiparallel (repulsive) acceleration components. Due to the isotropy of small scales $<10\eta$ we interpret this difference as a result of the effective electrostatic forces. The dashed line is a fit $\sim 1/r^2$ from which we estimate the charge on the particles.}
    \label{fig:InteractionForces}
\end{figure*}

The particles are injected into the wind tunnel flow using an in-house particle dispenser depicted in Fig. \ref{Fig:ParticleDisperser}. It consists of a reservoir (blue) and a smaller cavity (green) connected to an external SF\textsubscript{6} supply and a nozzle through flexible metallic pipes. For each experiment, a portion of the particles falls into the cavity and is washed away by a flow of SF\textsubscript{6} towards the nozzle, where the fluid shear forces break up agglomerates to form a cloud of mostly mono-disperse particles (see Fig. \ref{Fig:ParticleSizePlot}). The number of particles within the cavity cannot be controlled precisely, which causes most of the variation in the seeding density apart from the wind tunnel turbulence itself.

The SF\textsubscript{6} supply is a conventional gas bottle with a pressure regulator. The pressure regulator is set to 2-5 bar above the tunnel's pressure, which is one means of influencing the seeding density in the measurement volume. Since the particle dispenser itself is not free of leaks, the gas supply is interrupted between experiments through a magnetically actuated valve. A second magnetically actuated valve at the outlet of the dispenser controls the flow of SF6 through the setup and the subsequent release of particles through the nozzle. Its opening time is the second means of controlling the seeding density. The nozzle is a 1.2~mm Laval nozzle removed from a commercial airbrush. All parts of the particle dispenser are carefully electrically grounded and the ground connection is verified whenever the setup is adjusted. 

The ejection nozzle is placed between flaps of the active grid 15cm left of the measurement section centerline at the upstream end of the measurement section. The nozzle points about 40 \degree upwards. This ensures that the seeded portion of the flow is dominated by the approximately homogeneous and isotropic active grid turbulence instead of the jet ejecting the particles. It furthermore homogenises the seeding density. 

Within the reservoir, two metallic cones (yellow) are connected to a stepper motor, which moves them vertically. This mechanism allows for remote declogging of the device, but its usage is very rarely required, likely due to the frequent short flows of SF\textsubscript{6}. We achieve seeding densities of up to 130 particles/cm\textsuperscript{3} and eject an estimated 0.1 ml of particles per recorded video. 

We have verified that the particle dispenser releases predominantly single particles. For this we have placed a microscope slide on the floor of the measurement section during experiments. The ejected particles settled onto this slide and their size distribution was measured using microscopic images and ImageJ-based particle sizing. Fig. \ref{Fig:ParticleSizePlot} (A) shows that predominantly single particles with a narrow size distribution were ejected. 

Pneumatic conveyance as implemented here is known to electrostatically charge the conveyed particles. To assess the effect of electrical charge of the particles on the flow measurement, we followed two approaches: First, we measured the radial distribution function (RDF) of the particles, i.e. the relative probability of finding a particle a distance $r$ away from another. The RDF of electrostatically repelling particles in a turbulent flow differs from the RDF of uncharged particles \cite{Lu2010,Lu2015} for distances $r$, where the electrostatic forces exceed the fluid forces. 
Second, we calculate the vectorial difference between the accelerations of two adjacent particles $\mathbf{\delta a}=\mathbf{a_1}-\mathbf{a_2}$ separated by $\mathbf{\delta r}=\mathbf{r_1}-\mathbf{r_2}$. If their interaction is predominantly of electrostatic nature, $\mathbf{\delta a}$ and $\mathbf{\delta r}$ are (anti)parallel. 
\begin{figure*}[t]
  \includegraphics[width=\textwidth]{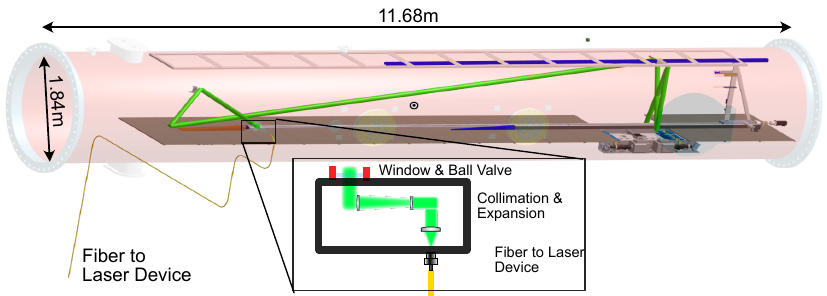}
  \caption{Illustration of illumination and imaging in measurement section of the wind tunnel. Green laser light from a Yb:YLF-laser is guided towards an optics box through an optical fiber. The diverging beam is collimated and expanded through lenses. It enters the tunnel as an approximately 3.5 cm wide beam with minimal divergence. The beam is guided towards the opposite end of the measurement section, where an arrangement of three mirrors forms an "X" with the measurement volume in its joint. The cameras reside below the measurement section floor and observe the measurement volume through optical windows.}
  \label{Fig:BeamPath}
\end{figure*}
We have performed the above analysis on every 25th frame of the dataset that shows the smallest mean acceleration values and is thus most likely to show charge biases. Increasing the number of frames did not qualitatively change the results. Fig. \ref{fig:InteractionForces} (A) shows histograms of the angle between $\mathbf{\delta a}$ and $\mathbf{\delta r}$. For large inter-particle distances we observe clear preference for a perpendicular alignment between $\mathbf{\delta a}$ and $\mathbf{\delta r}$ (as expected for fluid forces), whereas a parallel alignment (expected for electrostatic interactions) does practically not occur. At inter-particle distances of  $<180\mu$m, the PDFs are shifted towards larger angles indicating that the particles accelerate away from each other. This is corroborated by Fig. \ref{fig:InteractionForces} (B), where we calculate the mean magnitude of the particle acceleration projected onto their separation line $\mathbf{r}$. While this value is expected to increase towards smaller separations in turbulent flows with uncharged particles \cite{Borgas1998}, there is no reason for the repulsive interactions to be more pronounced than the attractive ones at those small and isotropic scales. We interpret the mismatch as the result of charges on the particles, which allows us to obtain a rough estimate of the electrostatic force between them. The inset in Fig. \ref{fig:InteractionForces} (B) shows this difference and a fit of $C/r^2$. By applying Coulomb's law we obtain an estimate for the charge on each particle ($\sim 10^4$ elementary charges) from $C$. 

The presence of charge effects at close distances is also seen in the radial distribution function (RDF). Fig. \ref{fig:ExampleMeasurements} shows that going from larger to smaller increments the RDF begins to decrease around 60 $\mu$m, which indicates the presence of a unipolar charge \cite{Chun2005,Lu2010}. Following \citet{Chun2005} we can estimate the charge on the particles from the position of the peak around 60 $\mu$m. We arrive at $10^4$ elementary charges in good agreement with the estimate presented above. 

\section{Illumination}

The illumination of the particles is provided by a frequency-doubled TruMicro 7240 Yb:YAG laser (Trumpf Laser GmbH, Schramberg, Germany) with a wavelength of 515~nm, and a maximum energy of 7.5~mJ per pulse. It releases pulses of 300~ns duration at 20-100~kHz yielding a maximum power of 300~W. The laser is commonly used for precision-welding applications, but has been used in the research context to illuminate cloud droplets\cite{Bertens2021}. It features a low-power alignment laser (Class 2, 630-680 nm), which makes the alignment of optics in the open experimental hall and inside the tunnel safe. The high-power Yb:YAG laser is only operated when the tunnel manholes are closed and secured from operation otherwise through adequate procedures for laser safety.

The laser beam is guided through a 30~m long optical fiber (LLK-D06, 100mm~mrad, Trumpf Laser GmbH) into a light-sealed box. The box enclosure contains a custom mount for the optical fiber connector on one side and a round hole on the opposite, tunnel-facing side. The bottom of the box consists of an optical breadboard. The box is mounted flush on the optical access flange of the wind tunnel and the remaining gaps are covered with laser safety fabric. Its support structures are fixed to the facility walls, such that the relative motion between the laser beam and the wind tunnel are minimised. The beam leaves the fiber at an divergence angle of $(73\pm4)$~mrad. We use a single, collimating lens and a beam-expander to generate an almost parallel, approximately 4cm wide beam. The remaining divergence angle is so small that no further beam-shaping optics needed to be placed inside the tunnel despite the approximately 12.5~m long beam path. This meant a substantial simplification of the optical setup, since the focal length of optics within the wind tunnel depends on the facility pressure.

The beam enters the tunnel through a pressure sealed window (Typ 76, METAGLAS GmbH). The window is protected by a ball valve, which automatically closes in case of a flow through the flange. The laser safety circuit is automatically opened in this case preventing the laser to shine into the closed ball valve.
The beam enters at an angle, such that the fraction of light reflected from the uncoated, thick window cannot be focused back into the fibre. 

\begin{figure}
  \includegraphics[width=\columnwidth]{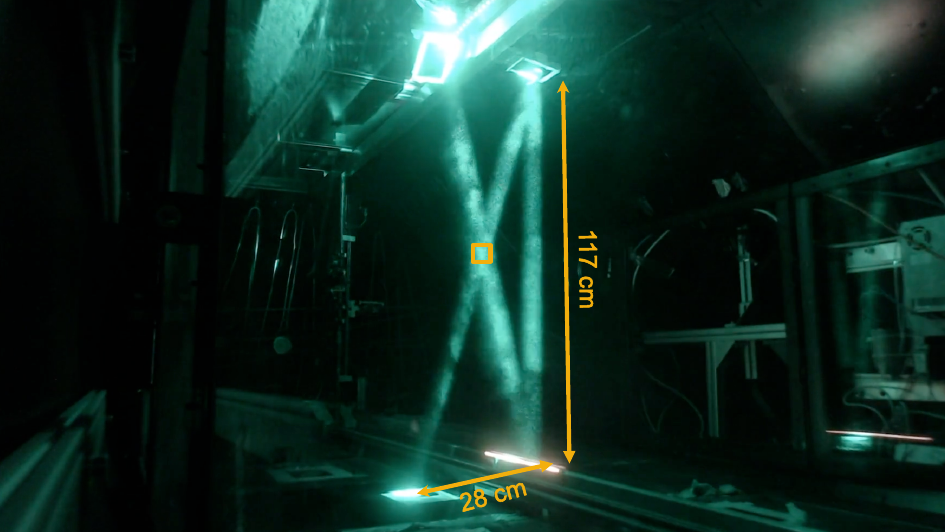}
    \caption{Self-crossing laser beam forming the measurement volume at its intersection. The laser beam first hits the upper left mirror and is dumped onto the steel floor in the lower left corner of the "X". Brighter spots in the beam are due to "clouds" of particles passing by. }
  \label{Fig:XPhoto}
\end{figure}
Inside the facility the beam is not enclosed further. The beam path within the facility is illustrated in Fig. \ref{Fig:BeamPath}. It first crosses the measurement section perpendicular to the flow direction towards a mirror. Its kinematic mount (AC-8823, Newport Optics) can be remotely adjusted by $\pm 3.5\deg$ from outside the facility, which is the only means of adjusting the beam path once the wind tunnel is pressurised. The beam is directed onto a 2-inch mirror fixed onto the tunnel floor using conventional lens posts. A 3D-printed aerodynamic housing has been manufactured for this mirror to reduce the flow disturbance, prevent misalignment by the flow, and reduce the cover of particles. 

The beam is then directed towards the imaging setup approximately 6.5~m downstream of the beam entry into the tunnel. Starting from the wind tunnel ceiling, an arrangement of fixed-angle mirrors guide the beam such that it forms an "X" parallel to the wind-tunnel cross section (see Fig \ref{Fig:BeamPath} and \ref{Fig:XPhoto}). The intersection of the "X" is a  double-cone with a maximum diameter of 4.5 cm. In this region, where the laser beam passes twice, the amount of light is sufficient to illuminate the 10 $\mu$m large particles to allow tracking even when they are slightly out of focus. The configuration has the additional advantage that all cameras experience similar scattering angles, even though they observe the measurement volume from four opposite directions (see Sec. \ref{Sec:Imaging}). 
	
The beam is dumped onto a black steel floor panel. The position of the beam on the mirrors can be observed through several webcams. In addition, we use a flat plate on a remote-controlled traverse (the opposite side of the calibration mask) to observe the position of the alignment laser. We mark the correct beam position at well-defined traverse positions while the tunnel is accessible for maintenance. When it is filled with SF\textsubscript{6}, it is straightforward to check the beam position using a standard webcam. We observe that the beam position moves by up to a centimeter, but only when changing the facility pressure. Please note, the tunnel  can deform by up to 0.5~cm under pressure, which in turn moves the mirrors mounted to the walls. This likely explains the observation. The shifts in the beam position were  corrected using the remote-controlled kinematic mirror. 

A train of laser pulses of pre-defined frequency and duration was released upon receiving a digital signal from the digital signal generator described in Sec. \ref{Sec:Imaging}. This was realised through the proprietary programming interface of the laser device. The laser frequency was set to the camera sampling frequency or a multiple thereof, the duration of the illumination was chosen as  $4-10~\times$ the video duration (0.56s). 

\section{Imaging and operating procedures} \label{Sec:Imaging}
\begin{figure*}[ht]
  \includegraphics[width=\textwidth]{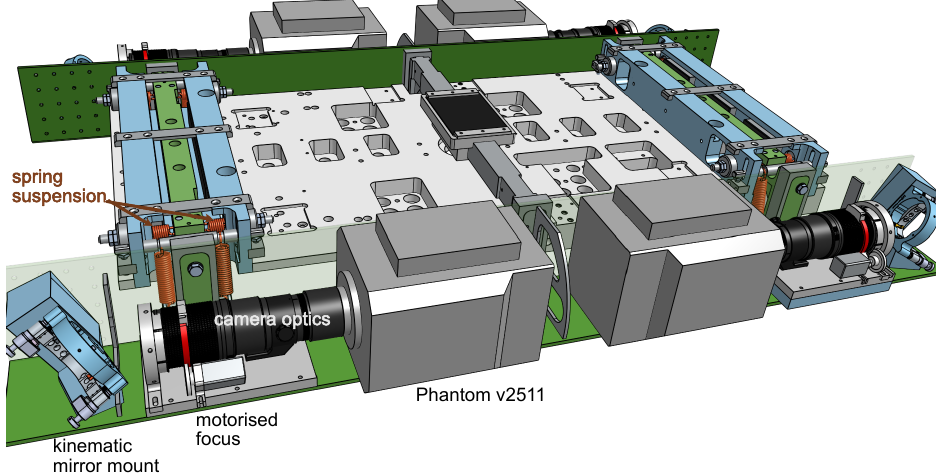}
  \caption{Camera platform located underneath the wind tunnel floor. The two camera support structures (green) are connected rigidly with each other, but connected to the rest of the platform through vibration/damping springs (red).  Each of the four high-speed camera observes the measurement volume through a 200mm telephoto lens, two $2\times$ teleconverters, a mirror, and an optical window. The mirrors can be remotely adjusted by motors.  The lenses are focused remotely through timing belts, which are moved by a stepper motor. The measurement volume is located approximately 60cm above the center of the traverse. }
\end{figure*}
Modern Lagrangian particle tracking relies on the ability of high-speed digital cameras to capture the positions of multiple particles sufficiently fast to allow their reliable frame-by-frame tracking in a vigorous turbulent flow. Full information can only be extracted in three-dimensional particle tracking, which requires at least two cameras. However, three cameras dramatically increase the fraction of trackable particles and four cameras allow the noise-free reconstruction of acceleration statistics \cite{Lawson2018,Mordant2004}. 

The particle tracking setup presented here allows for the simultaneous recording of up to four Phantom v2511 high-speed cameras. These cameras have a maximum recording speed of 25.6 Gpx/s (25 kHz at full resolution of $1280 \times 800$ px). They have been customised by the manufacturer to allow their operation under varying external pressure. Specifically, the protective glass on top of the sensor was vented to allow for pressure equilibration. Otherwise, as most electronic equipment, the cameras can be operated in pressurised SF\textsubscript{6} without further precautions. 
The camera recording frequency is synchronised with the laser pulse frequency. Recordings are triggered by the digital output interface described below. 

The camera optics consisted of a AF Micro-Nikkor 200 mm 1:4D IF-ED camera lens, whose aperture is set to $f/11$, and two $2\times$ teleconverters resulting in a magnification close to 1 when focused to the center of the measurement volume. Since the refractive index of SF\textsubscript{6} depends sensitively on the pressure \cite{Thomas1988}, the cameras need to be refocused after changing pressure. We mounted a timing belt to the manual focus rings of each 200mm camera lenses and connected them mechanically to stepper motors (Trinamic QSH2818-51). The motors were controlled remotely through a Arduino Motor Shield using a MATLAB program as external interface. 

The cameras and imaging optics were mounted on spring-suspended platforms. The  springs connected the platforms to a sled, which rests on the tunnel rail system \cite{Bodenschatz2014a}. This arrangement effectively decoupled the imaging setup from the tunnel structure, which vibrated during operation. Remaining motions of the cameras and imaging optics were eliminated by the dynamic camera calibration of the particle tracking code.

To extract real-world physical measurements from the sensor coordinates, the cameras need to be calibrated. That is, the sensor coordinates on each camera need to be related to coordinates in the three-dimensional measurement volume. A standard method to perform such a camera calibration is the placement of a flat plate with known features (e.g. dots or a square pattern) at different positions within the measurement volume. The knowledge of the exact calibration plate position is not strictly necessary \cite{Muller2020,Zhang2000} if the angle of the calibration plate with respect to the sensor plane can be changed significantly. In an enclosed environment it is easier to linearly traverse a calibration plate through the measurement volume and record the traverse positions. Since the cameras are refocused at each measurement pressure, their camera model needed to be updated in-situ and the calibration had to be remotely controllable.  We thus mounted the calibration plate on a 500mm linear stage (igus GmbH), which was mounted on an existing horizontal instrumentation traverse. The calibration plate was a printed grid of 1mm diameter black circles with 2mm grid spacing glued onto a flat aluminium plate. Three of these circles were marked as fiducial markers. The calibration plate was illuminated by an LED strip.

A typical calibration procedure consisted of moving the horizontal traverse such that the calibration plate was visible on the high-speed cameras, traversing the plate through the measurement volume in steps of 2mm (traverse reading) and returning the instrument traverse to its original position at the downstream end of the wind tunnel section. Small motions in the position of the traverse and imperfections in the traverse mechanics and position reading made this calibration relatively imprecise. It was therefore only used as a starting point and the necessary precision was achieved by the on-track calibration and dynamic self-calibration of the tracking code (see Sec. \ref{Sec:TrackingCode} and \citet{Bertens2021}). 

 The camera clocks controlling the exposure of a single frame were linked together to the laser master clock. The video download and camera setup occured over 1GbE or 10GbE connections. The cameras were connected to network switches located inside the wind tunnel. A conventional ethernet cable (1 GbE) and three pairs of optical fibres connect these switches to the outside of the facility through electrical and optical feedthroughs, respectively. The download rate over the 10GbE connection varied between 2 and 8 Gb/s, such that the acquisition of a single experiment (96 GB in total) with four cameras takes between 4 and 10 minutes. 
 
 The data acquisition was controlled by TTL-signals over coaxial cables. The magnetically actuated valves, the proprietary laser interface, and the camera triggers were connected to a National Instruments USB-6229 Digital I/O USB-interface. This interface was controlled by a custom MATLAB-program, which sent signals to the valves, laser, and cameras in a pre-defined sequence. Specifically, the magnetically actuated valve 1 was opened first to pressurise the particle dispenser. After 1.5~s valve 2 was opened for a user-defined time to release a puff of particles. Valves 1 and 2 were then closed and the program waited for a user-defined time to release laser pulses. After a third user-defined time, the program sent a trigger signal to the cameras and they started their acquisition. This process could  be started automatically once the videos had been downloaded, i.e. several experiments could be carried out successively without supervision. 

\section{Tracking Code}\label{Sec:TrackingCode}

\begin{figure*}
    \centering
    \includegraphics[width=1.6\columnwidth]{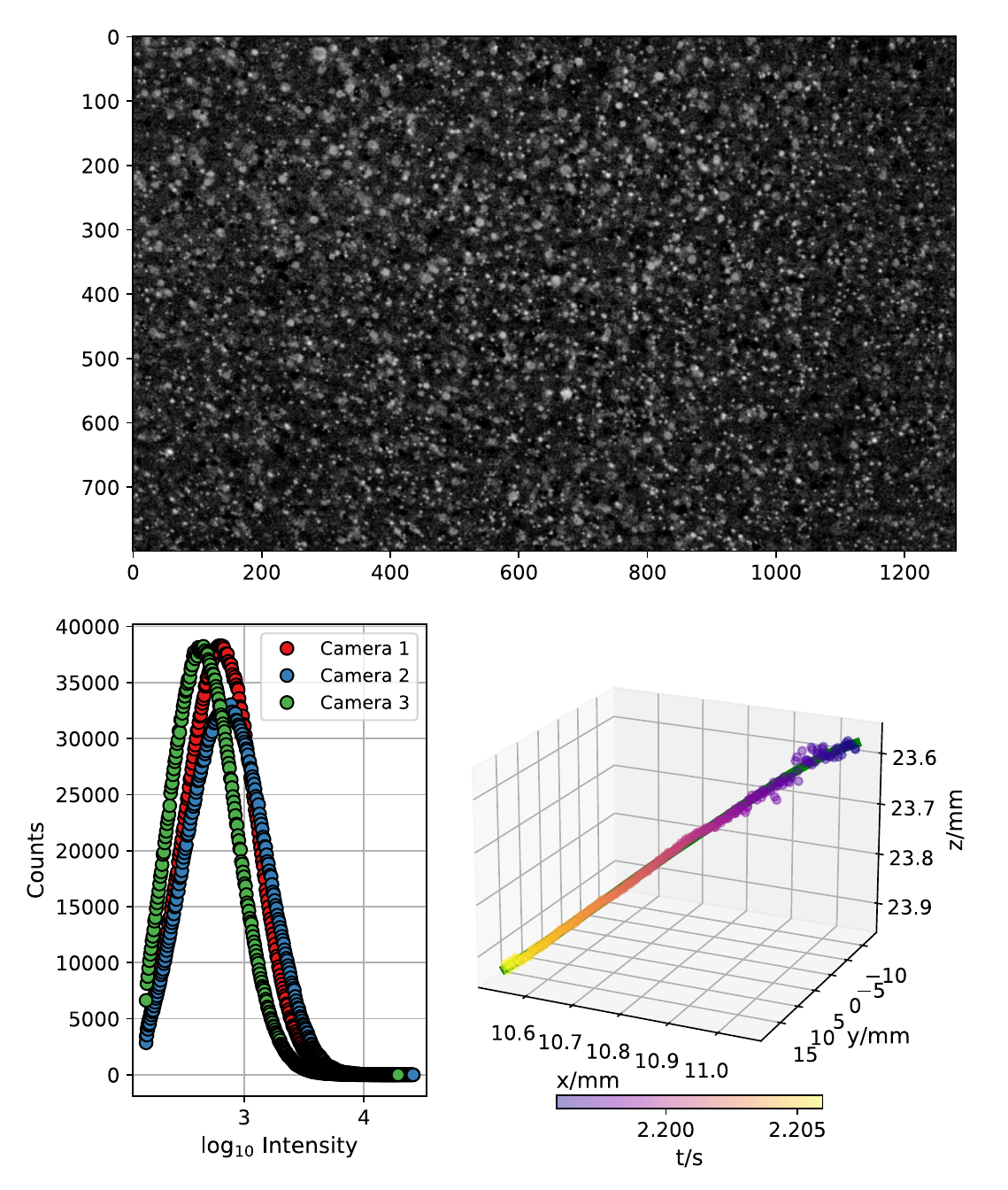}
    \caption{Particle tracking properties. \textit{Top:} Sample frame with a relatively large seeding density on Camera 2. Note the diffraction patterns on images of out-of-focus particles, which we use to inform the particle triangulation about the sensor-particle distance. \textit{Left:} Histograms of the logarithm of the particle intensity. \textit{Right:} Single sample track with raw data color-coded in time and the resulting smoothed track in green.}
    \label{fig:SampleFrame}
\end{figure*}

Fig. \ref{fig:SampleFrame} shows a sample frame from a typical video taken under the conditions specified in Tab. \ref{tab:TurbulenceParameters}. To track the cellulose particles we used the particle tracking algorithm originally developed for in-situ tracking of cloud droplets\cite{Bertens2021}. 
Many of the experimental challenges faced there,
such as insufficient illumination or large sweeping flow,
are present here to an even larger degree.
In particular, the necessarily small size of the particles resulted in
a very low amount of light scattered by them onto the camera sensors,
even with relatively large aperture diameters ($f/11$). Fig. \ref{fig:SampleFrame} shows an example of the intensity distribution on a typical video. 
This lead to low signal-to-noise ratio and less position accuracy of the images.
On the other hand, unlike the cloud droplets the cellulose particles are always small enough so that they can be considered as point sources of light, making our choice of point spread function universally valid.

The thermal gradients generated by the camera fans,
combined with the high dependence of index of refraction on temperature
for SF6 at high pressure, significantly lower the image quality
and shift the apparent centres of the particle images in a manner
that cannot be accommodated by camera model self-calibration.
These apparent shifts preclude usage of standard shake-the-box algorithm, and justify our changes to it as described below.

The detailed description of our particle tracking algorithm warrants
a separate publication, and its main features have been described in \citet{Bertens2021}
Here we briefly summarise the parts that are crucial
to successful tracking in our wind tunnel.

The tracking algorithm is broadly based on the shake-the-box (STB) algorithm developed by \citet{Schanz2016}.

Like the standard STB, we use the progressive subtraction of
already-tracked objects and iterative improvements of their fitted parameters. Unlike the standard STB, we do not tie the particle image locations to the particle three-dimensional positions (as the thermal gradients make the link unreliable), but only use the projected image positions as starting guesses for the iterative optimisation process, which happens entirely in two dimensions.
We also use a more sophisticated stereoscopic reconstruction process, which takes into account not just image locations, but also their brightness and defocus.
The temporal linkage is delayed by several frames relative to the current frame (we chose to delay by six frames at 25 kHz), which is achieved by making each trajectory consist of potentially several heads (which temporally link the most recent positions)
and a single tail. The tail is extended by choosing its most likely continuation from the backs of all heads, after which the heads not connecting to the extended tail are pruned. This delayed decision process allows us to deal with the short residence time of the particles in our stationary setup, and to allow reliable temporal linkage even in the presence of strong accelerations. For more detailed description of the tracking algorithm, see \citet{Bertens2021}.

\section{Postprocessing of Raw Tracks}
The overall aim of the setup is to measure particle velocity and acceleration in the Lagrangian sense. To characterise the flow - in particular the rate of dissipation of turbulent kinetic energy - access to Eulerian quantities is desirable as well. The raw tracks contain random noise, systematic instrumentation-induced errors, and statistical sampling biases due to the time-dependence of the particle seeding. In the following we describe the processing steps implemented to reduce these biases:

\paragraph{Random Noise} To remove random noise from the trajectories, three methods are commonly used in the literature: Smoothing by convolution with (differentiating) Gaussian kernels is a frequently-used technique (e.g. \citet{Mordant2004}) and is similar to a conventional window smoother. However, the filter operation is undefined at the edges, which leads to a loss of data and a selection bias in the resulting statistics \cite{Lawson2018}. The first Lagrangian measurements at high Reynolds numbers employed Savitzky-Golay-filter (polynomial fits through portions of a track) to remove random noise from the raw trajectories \cite{Voth2002}. However, the resulting statistics are very sensitive to the choice of the filter length (the length of portions) and the filter is undefined at the edges of a track, although the effect is smaller than in the Gaussian filter \cite{Voth2002}. B-Spline filtering \cite{Eilers1996} was made popular for smoothing Lagrangian particle tracks in turbulent flows by \citet{Gesemann2016} and compared to Gaussian filtering by \citet{Lawson2018}. 

\begin{figure}
    \centering
    \includegraphics[width=\columnwidth]{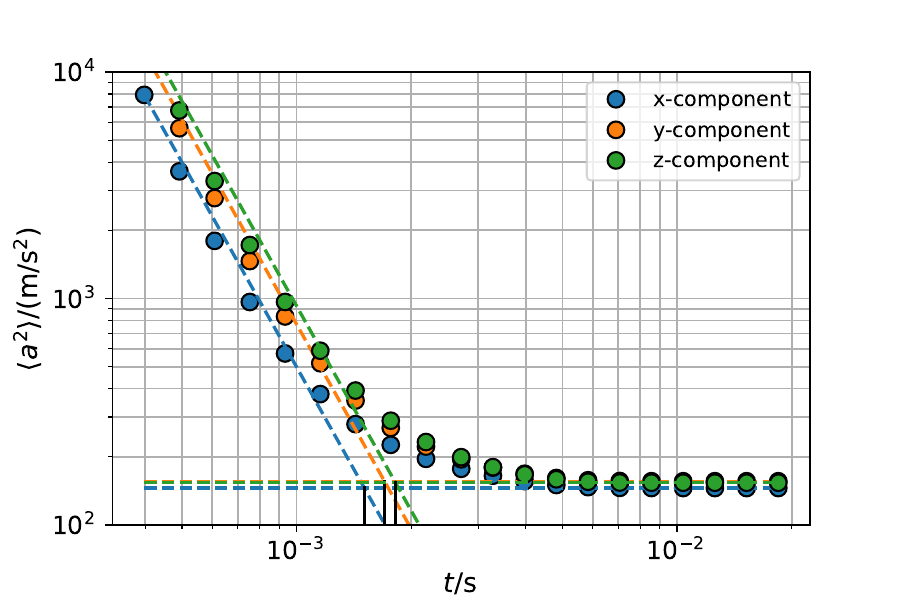}
    \caption{Procedure to find the ideal smoothing length. At small $t_f$, noise dominates the acceleration variance yielding too high values. Since the noise is approximately white, $\langle a^2 \rangle \sim t_f^{-{1/2}}$ (falling dashed lines). For large $t_f$, $\langle a^2 \rangle$ approaches a constant (horizontal dashed lines) indicating oversmoothing. We choose the intersection of these two lines as ideal filtering lengths (solid black lines).}
    \label{fig:Smoothing}
\end{figure}

We have implemented a version of the B-Spline algorithm after \citet{Eilers1996}. We use the tracking code's internal measure of triangulation quality, namely the ratio of individual and typical triangulation error (sum of square distances between projection of best-fit particle position and particle image over all cameras), to construct a weighting matrix.
This reduces the filtering time scale by more than 50\%. 
To obtain an optimal filtering length $t_f$, we calculate the acceleration variance for a range of $t_f$ and apply the following procedures illustrated in Fig. \ref{fig:Smoothing}. For choices of $t_f$, where noise dominates the acceleration statistics and the smoothing is insufficient, $\langle a^2 \rangle \sim t^{-{1/2}}$. For large values for $t_f$, $\langle a^2 \rangle$ approaches a constant. In this latter regime the tracks are oversmoothed and the reconstructed trajectories do not resemble the data appropriately. We therefore choose the intersection of the $\langle a^2 \rangle \sim t_f^{-{1/2}}$-line and the $\langle a^2 \rangle = \mathrm{const.}$ as "ideal" smoothing parameter. Fig. \ref{fig:SampleFrame} shows an example result of the smoothing procedure. 

\begin{figure}[h!]
	\includegraphics[width=\columnwidth]{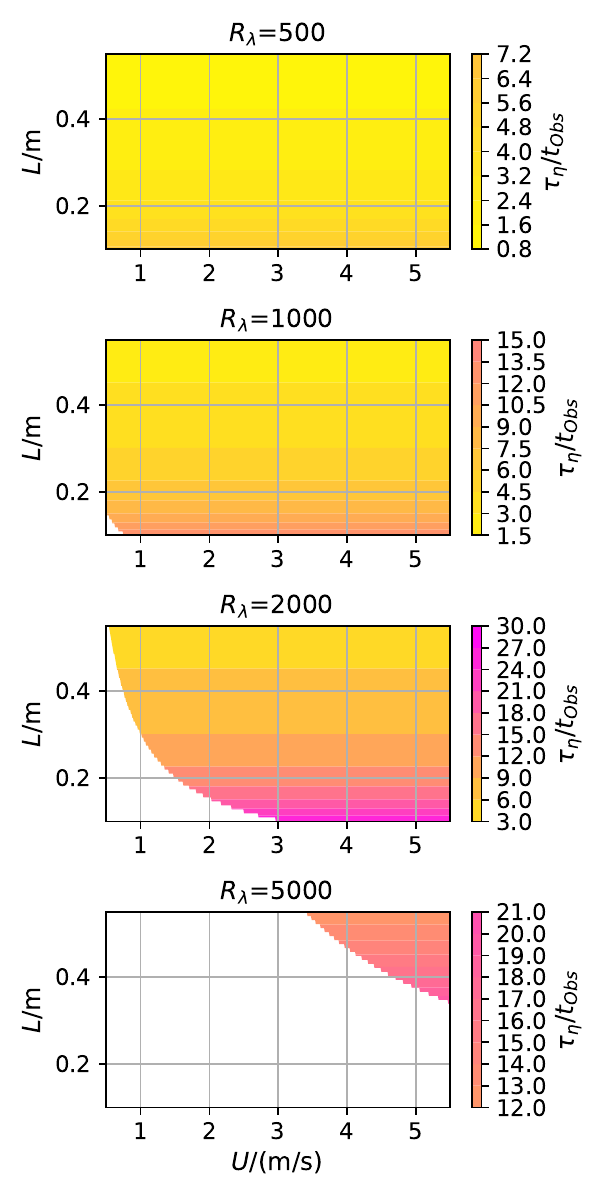}
	\caption{Number of $\tau_{\eta}$ for which a particle remains observable at different $R_\lambda$ given a turbulence intensity of $u_{\text{RMS}}/U=0.1$, as a function of $U$ and $L$. A given $R_\lambda$ can be obtained in multiple ways by adjusting the viscosity through the facility pressure, $L$ through the active grid and $U$ using the wind tunnel fan. Smaller $L$ and smaller viscosity allow a larger number of $\tau_\eta$ to be observed before the particles leave the measurement volume. White areas indicate parameter combinations inaccessible to the facility. }
	\label{Fig:ParameterStudy}
\end{figure}
\paragraph{Inhomogeneous Particle Seeding} The amount of particles per frame is critical for the statistical and spatial resolution of Eulerian quantities. The active grid causes the turbulence intensity to vary strongly from eddy to eddy. Depending on the mean flow speed and active grid agitation, the particle tracking setup "sees" between 0.5 and 5 large eddies of size $L$ per video. In previous measurements with nanoscale hot wires \cite{Vallikivi2011,Kuchler2019} averaging over several hours was a straightforward strategy to achieve well-resolved statistics. While more videos can be taken, the statistical convergence obtained with single-point hot wire measurements is out of reach. Moreover, the particle seeding throughout one video varies massively, such that some parts of the bypassing flow contribute disproportionately towards the ensemble statistics. Since even second-order flow statistics, such as the turbulence dissipation rate $\varepsilon$ have a heavy-tailed distribution, this can cause severe biases at all scales. 

The combined unsteadiness of the flow and its seeding must be considered when calculating Eulerian quantities, e.g. to obtain the rate of dissipation. We refer to the following section for an example of a potential strategy to mitigate these biases. 

\section{Characterisation}
In the following we analyse a representative measurement taken with the setup described here. The turbulence parameters can be found in Table \ref{tab:TurbulenceParameters}. 

\begin{table}[]
    \centering
    \begin{tabular}{c|c|c|c|c|c|c|c}
        $R_\lambda$ & $\varepsilon$/(m$^2$/s$^3$) & $P$ & $U$ & $u_{RMS}$ & $\tau_\eta$ & $\eta$ & St\\
                  2100 & 0.1 & 6 bar & 4.16 m/s & 0.42 m/s & 2.0 ms & 28 $\mu$m & 0.16
    \end{tabular}
    \caption{Parameters of the sample measurement chosen to characterise the measurement}
    \label{tab:TurbulenceParameters}
\end{table}

\paragraph{Parameter Ranges}
To characterise the data acquired from the setup we begin by studying the range of accessible parameters theoretically. Based on the range of mean flow speeds (0.5-5 m/s), energy injection scales (0.1-0.5~m), and a typical turbulence intensity of 10\% we use $\varepsilon=u_{RMS}^3/L$ to estimate the Reynolds number and Kolmogorov time scales $\tau_\eta$ accessible in the experiment. $u_{RMS}$ denotes the RMS of the fluctuating velocity component. We further estimate the maximum track lengths that can be expected based on the length of the measurement volume in the mean flow direction (4~cm). Fig. \ref{Fig:ParameterStudy} shows that low integral length scales and high pressures are advantageous if long track lengths are required, whereas a high relative temporal resolution can be obtained at large values of $L$ and small pressures. The parameter space at the largest Reynolds numbers is naturally limited to the most extreme parameters possible in the facility. While the choice of mean flow velocity $U$ appears to be unimportant at a given turbulence intensity, a subtle effect is not captured in this illustration: Thermal plumes rising from the warm high-speed cameras considerably impact the video quality at larger pressures ($>~6~\mathrm{bar}$) due to the strong temperature-dependence of the refractive index \cite{Arnaud1979,Thomas1988}. Higher mean flow speeds help improving the video quality, since they advect these thermal plumes away from the cameras.

We emphasise that the facility is capable of generating one $R_\lambda$ with a variety of parameter combinations. In particular, effects of the particles' finite Stokes number can be investigated by performing multiple experiments at a single $R_\lambda$, but different $\tau_\eta$. Since larger cellulose particles are available from the manufacturer, this furthermore allows the systematic study of inertial particle dynamics in turbulence over a wide range of Reynolds and Stokes numbers. Fig. \ref{Fig:ParticleSizePlot} shows that the Stokes number is below 0.1 (and the particles therefore good tracers) for Reynolds numbers up to 3000. By changing the experimental parameters, particle dynamics up to $\mathrm{St}\approx 1$ can be measured at the same Reynolds number. This opens for example the possibility for studies on the effect of particle size on clustering dynamics at atmospheric Reynolds numbers, which is of great interest to studies of cloud formation. 

\begin{figure*}
    \centering
    \includegraphics[width=\textwidth]{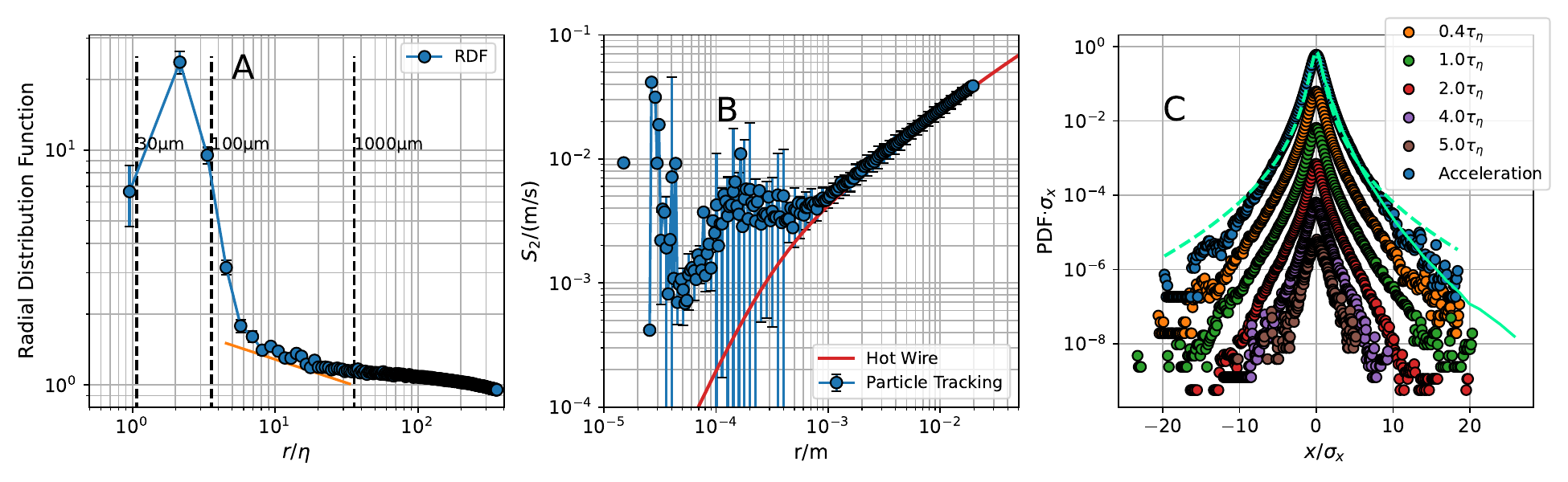}
    \caption{Turbulence statistics from an experiment with $R_\lambda\approx 2100$ (see Tab. \ref{tab:TurbulenceParameters} for details) (A): Radial Distribution Function (RDF) of a representative experiment at $R_\lambda\approx2100$(see Table \ref{tab:TurbulenceParameters}) as a function of the inter-particle distance $r$ normalised by the Kolmogorov length scale $\eta$. The orange line indicates the power law expected for St=0.2 with an exponent of -0.2. The dotted lines indicate distances of 30 $\mu$m, 100 $\mu$m, and 1 mm. (B): Second-order longitudinal Eulerian structure function $S_2 = \langle((\mathbf{u}(\mathbf{x}) - \mathbf{u}(\mathbf{x}+\mathbf{r}))\cdot \frac{\mathbf{r}}{|\mathbf{r}|})^2\rangle$ measured from particle tracks (blue circles) and nanoscale hot wires (red line) \cite{Kuchler2020}. (C) PDFs of the velocity increments along the particle tracks and particle acceleration normalised by their respective standard deviations. Curves have been shifted by one decade for better visibility. Green dotted line is the stretched exponential typically found in the acceleration PDFs of tracer particles \citet{Voth2002,Mordant2004}. The green dotted line is the PDF from DNS \citet{Bec2006} at St=0.16, which is close to the Stokes number found here. }
    \label{fig:ExampleMeasurements}
\end{figure*}

\paragraph{Radial Distribution Function}
An important question in the study of particles in a turbulent environment is how they distribute spatially  within a flow. A perfect tracer will homogeneously seed the flow, but particles with a finite inertia will be expelled from regions with strong vorticity leading to an undersampling of those regions and a clustering of particles in other regions (e.g. \citet{Gibert2012}). Such clustering effects are captured by the relative probability of finding a particle at a distance $r$ away from a different particle, i.e. the radial distribution function (RDF). For perfect tracers the RDF is independent of $r$, whereas clustering leads to an increase of the RDF at small $r$. A simple way of obtaining the RDF is to calculate the probability of finding a certain $r$ within a dataset and compare it to the probability of finding a certain $r$ in an artificial frame containing a random sample of particles from the dataset. This latter step accounts for the reduced detection frequency of points close to the boundary or other setup-specific biases. A detailed account of this procedure is given in \citet{Saw}. Fig. \ref{fig:ExampleMeasurements} (A) shows the RDF for a flow at $R_\lambda \approx 2700$ measured at a pressure of 6 bar. We can identify five regions in this plot going from large separations towards small: (I) the drop at large scales $>100\eta$, which can be attributed to residual large-scale inhomogeneities due to incomplete mixing \cite{Saw2012a}, (II) the approximately flat region indicating a uniform distribution of particles, (III) the initial clustering due to finite stokes numbers, (IV) the strong clustering at scales $<4\eta$, and the drop of the RDF at very small scales $<60\mu$m, which is likely caused by charges on the particles as detailed in Sec. \ref{sec:Particles}. Region (III) can be approximated by a power law with an exponent of 0.2, which indicates a Stokes number of 0.2 in good agreement with the parametric estimate for this dataset. Region (IV) is reminiscent of a recent experimental result \citet{Hammond2021}, where the RDF shows a strong increase going towards very small separations. This feature requires an in-depth investigation beyond the scope of this paper. 

\paragraph{Eulerian Structure Functions}
The statistics of spatial velocity increments are a ubiquitous quantity in the study of turbulence at small scales and therefore a prime quantity to assess the capability of the setup to measure velocity statistics. Here we study the so-called longitudinal structure functions defined as $S_2 = \langle((\mathbf{u}(\mathbf{x}) - \mathbf{u}(\mathbf{x}+\mathbf{r}))\cdot \frac{\mathbf{r}}{|\mathbf{r}|})^2\rangle = \langle du(r)^2 \rangle$, i.e. the variance of the velocity difference at two points separated by $r$ projected onto $r$. It is accepted that to a reasonable approximation $S_2\sim r^2$ at small increments, $S_2\sim r^{2/3}$ in the inertial range and $S_2=u_{RMS}^2/2$ at large scales in the limit of large $r=|\mathbf{r}|$. In practice, we calculate $du(r)^2/du(r_{\mathrm{max}})^2$ for the particle pairs that we find within a frame, where $r_{\mathrm{max}}$ is the value in the largest increment bin, and calculate the binned mean for each frame. The per-frame normalised structure functions are then averaged and rescaled by $\langle du(r_{\mathrm{max}})^2 \rangle$. This procedure is inspired by \citet{Viggiano2021} and accounts for the different large-scale statistics and seeding densities passing the measurement volume during a video.  Fig. \ref{fig:ExampleMeasurements} (B) compares the particle tracking structure function to the one measured by nanoscale hot wires at an earlier point using the same experimental parameters \cite{Kuchler2020}. We see a very good agreement at scales $>1$mm, but strong discrepancies at smaller scales and large statistical uncertainties. These may be caused by several effects. First, the number of frames containing data at small increments is relatively small (1-100) resulting in poor statistical convergence. Furthermore, at small increments with poor statistics a precise measurement of the velocity increment is crucial and an error of just 0.03 m/s in the velocity measurement leads to an error of 300\% in the individual measurement of $S_2$. In hot wire measurements these errors are mitigated by large statistics and calibration errors cancel partially when calculating differences. Here, single velocity measurements with uncorrelated errors are subtracted leading to large errors in the resulting structure functions. Furthermore, the data begins to diverge around 200$\mu$m, where charge effects begin to play a role and might have an impact on the velocity difference statistics. 

\paragraph{Acceleration and Lagrangian Velocity Increments}
The acceleration of (tracer) particles in a turbulent flow is of great fundamental importance, since it is closely related to strong dissipative processes \cite{LaPorta2001,Voth1998,Voth2002,Ayyalasomayajula2006,Buaria2022}. It is known to be a highly intermittent quantity with heavy-tailed PDFs\cite{LaPorta2001,Mordant2004}. We show measurements of the accelerations and the velocity increments along the particle trajectories in Fig. \ref{fig:ExampleMeasurements} C. The prominent strong tails can be seen in all increments studied and get progressively weaker with increasing separation as expected. For the acceleration, \citet{Voth2002} suggested a stretched exponential that has shown to accurately describe acceleration statistics over a wide range of experiments and numerical simulations \cite{Mordant2004,LaPorta2001}. In Fig. \ref{fig:ExampleMeasurements} C we show the stretched exponential in green with parameters obtained by \citet{Mordant2004}, which lies above the data in the tails. The data is however in good agreement with data from numerical simulations\cite{Bec2006} with a Stokes number of 0.16. This Stokes number is consistent with St calculated from the parameters of this experiment as well as the power law exponent found in the RDF. 
The PDFs of finite velocity increments show thinner tails as we increase the increment up to $5\tau_{\eta}$, which is the largest increment where such PDFs can be drawn sensibly given the length of the particle tracks\cite{Lawson2018}. 

\section{Discussion}
Here we demonstrate an experimental setup capable of tracking particles in a turbulent flow at unprecedented $R_\lambda$. 
This widens the parameter space for Lagrangian measurements by about a factor of five compared to the previous state of the art. Lagrangian measurements have proven to be notoriously challenging and only been available for little more than 20 years. This explains the large leap in Reynolds number that was possible using a single experiment.
We have demonstrated that measurements of particle position, Eulerian and Lagrangian velocity increment statistics and Lagrangian accelerations are possible with this setup. 
We have introduced cellulose particles as a seeding material for flow measurements, where small spatial ($\sim 10\mu$m) and temporal scales ($\sim$ms) have to be resolved and humans come into extensive contact with the material. The material might thus be considered as an alternative to visualise flows beyond the laboratory environment, e.g. for studies of thermal comfort or airborne disease transmission \cite{Bourouiba2021,Hejazi2022,Bourrianne2021}.
The illumination intensity currently limits the particle size to $\geq10\mu$m, but smaller particles are available. 
To characterise the behaviour of these particles as a seeding material we have measured their density and found their density to be about half that of solid cellulose indicating the presence of voids in the material. Despite the decreased density, the particles cannot be seen as ideal tracers for a wide range of experimental parameters (see Fig. \ref{Fig:ParticleSizePlot} (D)). However, the three independently adjustable flow parameters (mean velocity, grid setting, pressure) allow effects of finite Stokes number to be systematically investigated while keeping the flow Reynolds number constant. At very large $R_\lambda>3000$ 10$\mu$m cellulobeads are inertial for all parameter choices. 

These new particles further required the development of a dedicated particle dispenser. We validated that the dispenser releases predominantly particle singlets.
We further measured the charge on the particles as they appear in the measurement volume in two independent ways yielding consistent values of about $10^4$ elementary charges. The electrostatic forces appear in the radial distribution function, where very small inter-particle distances appear to be depleted. However, $<0.5$\% of particles come closer than 200$\mu$m to another particle, such that electrostatic forces only need to be taken into account when such small distances are explicitly of interest. 
The particle dispensers's reliability is an area of potential future improvement. For example, the number of particles released varies considerably between injections. While no measurement campaign had to be interrupted so far because of a failed particle release, the dispenser needs to be flushed regularly with a strong stream of SF\textsubscript{6} to loosen stuck material. 

A serious challenge for the setup is the presence of thermal plumes, which can drastically distort the image at low flow velocities and high pressures. Their mitigation should be a priority in a next round of improvements of the setup. 

Currently, the track length of the particles is limited to about 10 ms, due to the rapid advection of particles out of the relatively small measurement volume by the mean flow. This limits the setup to the measurement of velocity increment statistics up to $\sim10\tau_\eta$, which is not enough for a proper study of the Lagrangian inertial range. The measurement volume also limits the range of scales, where Eulerian statistics can be computed to no more than $4000\eta$. Enlarging the measurement volume would thus be another fruitful improvement of the setup, albeit a rather complex one. 

In its current form, the setup can be used to measure acceleration and particle clustering statistics over a wide range of Reynolds- and Stokes numbers. Lagrangian velocity increment statistics can be obtained as long as only short increments are of relevance.
While the focus of this work is on Lagrangian turbulence, the setup allows measurements of Eulerian quantities previously inaccessible within the facility. This includes transverse structure functions for increments $\gtrsim 1$mm and transverse components of single-point statistics, both of which are important to fully characterise the anisotropy in the flow. At large particle seeding densities, methods like Vortex-in-cell \cite{Schneiders2016} or physics-informed neural networks \cite{Cai2021} might be able to estimate the entire three-dimensional velocity field from the scattered data. 

The setup is sufficiently general for most of its current limitations to be lifted in the future. For example, a different illumination source with longer pulse widths might make smaller and more tracer-like particles feasible, since currently only 1\% of the camera's exposure time are illuminated by the laser. An additional cooling system for the cameras would reduce the distortions due to thermal plumes. Finally, the camera setup is already connected to rails on ball bearings and is therefore prepared to be moved along the flow allowing tracking for much longer times. However, the flow properties and the maximum video length limit the sensible range of motion to about 2m. Still, this would for example allow for measurements of two-particle dispersion at high $R_\lambda$, which is of great fundamental and practical interest.


%
%

%

\begin{acknowledgments}
The Variable Density Turbulence Tunnel is operated and maintained by A. Kubitzek, A. Kopp, M. Meyer, and A. Renner. We thank A. Renner for continuous technical support throughout the project. We thank G. Bertens, D. Fliegner, and H. Degering for help with networking and high-performance computing. The machine shop led by U. Schminke manufactured the camera platform and the particle disperser. We thank Freja Nordsiek for helpful discussions regarding particle choices and their interaction when charged. We thank G. Bertens for providing optics and networking expertise as well as helpful discussions. We thank B. Hejazi for help with data acquisition. We thank H. Xu, A. Pumir, N. Ouellette, and G. Voth for helpful discussions. 
\end{acknowledgments}

\bibliography{main.bib}

\vfill\eject
\section*{Appendix A: Calculation of particle charges from the RDF}
According to \citet{Lu2010}, particle charge causes a drop in the radial distribution function towards smaller distances $r$, i.e. the RDF peaks at $r=r_*$, where Coulomb- and inertial forces have similar magnitudes. They infer that 
\begin{equation}
    r_*^3=\frac{2kq^2\tau_\eta \tau_p}{m_p B_{nl} c_1},
\end{equation}
where $m_p$ is the mass of the particle, $q$ is its charge and $k=1/4\pi \epsilon_0$ is the Coulomb constant. $B_{nl}$ (taken as 0.09 following \citet{Chun2005}) is a dimensionless coefficient describing the particle pair diffusion and $c_1=0.2$ is the exponent of the RDF power law e.g. as an orange line in Fig. \ref{fig:ExampleMeasurements} (A). Rearranging for $q$ and reading off $r_*\approx 60~\mu$m, we arrive at $\approx 10^4$ elementary charges.

\end{document}